\documentstyle[preprint,aps,prb]{revtex}

\tightenlines

\begin{document}
\draft
\preprint{F/I/S, Nov. 1998}

\title {Spin current in
ferromagnet/insulator/superconductor junctions }

\author{S. Kashiwaya}

\address{Ginzton Laboratory, Stanford University,
Stanford, CA, 94305-4085, USA.\\
Electrotechnical Laboratory, Umezono, Tsukuba, Ibaraki
305-8568, Japan.\\
CREST, Japan Science and Technology Corporation (JST).
}

\author{Y. Tanaka and N. Yoshida}

\address{Department of Applied Physics, Nagoya University,
464-8603,
Nagoya, Japan. }

\author{M. R. Beasley}
\address{
Ginzton Laboratory, Stanford University,
Stanford, CA, 94305-4085, USA.}

\date{\today}
\maketitle
\begin{abstract}
A theory of spin polarized tunneling spectroscopy
based on a scattering theory
is given for tunneling junctions between
ferromagnets and $d$-wave superconductors.
The spin filtering effect of an exchange
field in the insulator is also treated.
We clarify that the  properties of
the Andreev reflection are largely modified due
to a presence of an exchange
field in the ferromagnets,
and consequently the Andreev reflected quasiparticle
shows an evanescent-wave behavior
depending on the injection angle of the quasiparticle.
Conductance formulas for the spin current
as well as the charge current are given as
a function of the applied voltage and the spin-polarization
in the ferromagnet
for arbitrary barrier heights.
It is shown that the surface bound states do not contribute
to the spin current and that the zero-bias conductance
peak expected for a $d$-wave superconductor
splits into two peaks under the influence of the
exchange interaction in the insulator.
\end{abstract}
\vspace{20pt}

\narrowtext
\section{INTRODUCTION}
\label{sec1}
The transport properties in hybrid structures between
ferromagnets and superconductors
have received considerable
theoretical and experimental attentions.
Interest in such structures includes spin-dependent
spectroscopy of superconductors and possible
device applications.
Since the Cooper pairs in spin singlet
superconductors are formed between
up and down spins,
the high density of spin injection
through a
tunneling barrier induces a spin imbalance.
This non-equilibrium state is expected to
result in a suppression of the critical temperature
and the critical current density
in the superconductor.
A large number of experimental studies on  spin-polarized
tunneling
have already been performed using conventional
metal superconductors such as Al and Nb
about 20 years ago \cite{Tedrow}.
However, the recent discovery of so-called colossal
magneto-resistance (CMR) in
Mn oxides compound has aroused new interest
in this field \cite{MnO1,MnO2}, because hybrid structure
fabrication of the spin-polarized
ferromagnets with
high-$T_{c}$ superconductors is now possible using these
materials
\cite{Goldman,Yeh}.
\par
On the other hand,
the properties of ferromagnet/insulator/superconductor
(F/I/S)
and ferromagnet/ferromagnetic-insulator/superconductor
(F/FI/S)
junctions have been
analyzed based on the assumption
that the conductance spectra correspond to the density
of states (DOS) of the superconductor weighted by the
spin polarization \cite{Tedrow,Sfilter1,Sfilter2}.
A theory for F/I/S junctions based on a scattering
method has been presented by
de Jong and Beeneker\cite{deJong}, and new aspects of
Andreev reflection have been revealed, and
also detailed comparisons between theory and
experiments have been accomplished \cite{Soulen,Upadhyay}.
However, these results are restricted to
isotropic $s$-wave superconductors.
\par
In contrast to $s$-wave superconductor cases,
at the interface of a $d_{{x}^{2}-{y}^{2}}$-wave
superconductor,
zero-energy states (ZES) are formed due to the interference
effect of
the internal phase of the pair potential \cite{Hu}.
Tunneling theory for $d_{{x}^{2}-{y}^{2}}$-wave
superconductors
has already been presented by
extending the BTK formula \cite{BTK} to include the
anisotropy
of the pair potential
\cite{Tanaka,Kashiwaya1,Kashiwaya2}.
The theory predicts the existence of zero-bias conductance
peak (ZBCP)
which reflects the formation of the surface bound states
on the $d$-wave superconductors.
In this paper, an exchange interaction is introduced on the
normal side of the junction and on the insulator
in order to analyze the spin polarized tunneling effects.
The bound-state condition and tunneling spectroscopy
of ferromagnet/$d$-wave superconductor junctions have
already been
analyzed in two papers \cite{Zhu,Zutic}.
They have revealed  several important features in charge
transport.
Here we will argue that the properties of
the Andreev reflection \cite{Andreev}
is largely modified due to the presence
of the exchange interaction.
In particular, the existence of an evanescent type of
the Andreev reflection, which is referred to as virtual
Andreev reflection (VAR), is explained
for the first time (see Ref. 27).
This process has significant roles on the
transports especially for
junctions between half-metallic ferromagnets
and superconductors.
The conductance formulas for the charge and the spin
currents
are presented based on the scattering method
by fully taking account of the VAR process.
The merit of a formula based on  the scattering methods
is that
the conductance spectra can easily be calculated for
arbitrary
barrier heights cases
without the restriction of  the high-barrier limit.
The spin current is, we believe,
the most important physical quantity
in spin injection devices
based on the following two reasons:
one is that the spin current  gives a direct criteria to
estimate the effect of the spin imbalance
induced by the tunneling current,
the other is that
the charge and the spin conductivity
may illuminate the study
of electron systems that undergo
 spin-charge separation,
such as Tomonaga-Luttinger liquids and
possibly underdoped
high-$T_{c}$ superconductors \cite{Zhao,Si,Fisher}.
We will also analyze ferromagnetic insulator effects,
which includes the spin-filtering effect
\cite{Sfilter1,Sfilter2},
due to the presence of an exchange field in the insulator.
It is shown that a spin-dependent energy shift
during the tunneling process induces a splitting of the
ZBCP.
Based on the detailed analysis of the conductance spectra,
we propose a simple method to
distinguish the broken time-reversal symmetry (BTRS)
states inducement at the surface
\cite{Matsu,Fogel,Covington}
from spin-dependent tunneling effects.
The  implications of the ferromagnetic insulator effects
on tunneling experiments of high-$T_{c}$ superconductors
and a proposal for possible device applications are also
presented.
\par
\section{Formulation}
\label{sec2}
For the model of formulation,
a planar $F/FI/S$ junction
with semi-infinite electrodes
in the clean limit is assumed.
A flat
interface is assumed to be located at $x=0$,
and the insulator for up [down] spin is described by a
potential
$V_{\uparrow[\downarrow]}(\mbox{\boldmath $x$})$
\{$V_{\uparrow[\downarrow]}(\mbox{\boldmath $x$}) =
({\hat V}_{0}-[+]{\hat U}_{B})\delta(x)$\}, where
$\delta(x)$,
${\hat V}_{0}$ and ${\hat U}_{B}$ are the $\delta$-function,
a genuine barrier amplitude and an exchange amplitude
in the barrier, respectively.
The effective mass
{\it m} in the ferromagnet and
in the superconductor are assumed to be equal.
For the model of the ferromagnet,
we adopt the Stoner model where
the effect of the spin polarization is described by
the one-electron Hamiltonian with an
exchange interaction
similarly to the case of Ref.~\cite{deJong,Zhu,Zutic}.
For the description of the $d_{{x}^{2}-{y}^{2}}$-wave
superconductor,
we apply the quasi-classical approximation
where the Fermi energy $E_F$ in the
superconductor is much larger than the pair potential
following the model by Bruder \cite{Bruder,Beasley}.
The effective Hamiltonian (Bogoliubov-de Gennes equation) is
give by
\begin{equation}
\label{BdG}
\left[
\begin{array}{cc}
H_{0}(\mbox{\boldmath $x$})-\rho U(\mbox{\boldmath $x$}) &
\Delta(\mbox{\boldmath $x$},\theta)\\
\Delta^{\ast}(\mbox{\boldmath $x$},\theta) &
-\{H_{0}(\mbox{\boldmath $x$})
+\rho U(\mbox{\boldmath $x$}) \}\\
\end{array}
\right]
\left[
\begin{array}{l}
u(\mbox{\boldmath $x$},\theta)\\
v(\mbox{\boldmath $x$},\theta)
\end{array}
\right]
=
E
\left[
\begin{array}{l}
u(\mbox{\boldmath $x$},\theta)\\
v(\mbox{\boldmath $x$},\theta)
\end{array}
\right]
\end{equation}
Here, $E$ is the energy of the quasiparticle,
$U(\mbox{\boldmath $x$})$ is the exchange potential
given by $U\Theta(-x)$ ($U \geq 0$) where $\Theta(x)$
is the Heaviside step function,
$\rho$ is 1 [-1] for up [down] spins,
$\Delta(\mbox{\boldmath $x$},\theta)$ is the pair potential
and $H_{0}(\mbox{\boldmath $x$})
\equiv -\hbar^{2}\nabla^{2}/2m+V(\mbox{\boldmath $x$})-
E_{F}$.
To describe the Fermi surface difference
in F and S, we assume
$E_{F}=E_{FN}$ for $x<0$ and
$E_{F}=E_{FS}$
for $x>0$.
The pair potential $\Delta(\mbox{\boldmath $x$},\theta)$ is
taken as
$\Delta(\theta) \Theta(x)$ for simplicity.
The number of up [down] spin electrons is described by
$N_{\uparrow}$
[$N_{\downarrow}$].
The polarization and the wave-vector of quasiparticles
in the ferromagnet for up [down] spin
are expressed as
$P_{\uparrow} \equiv
\frac{N_{\uparrow}}{N_{\uparrow}+N_{\downarrow}}
=\frac{E_{FN} + U}{2E_{FN}}$
[$P_{\downarrow} \equiv \frac{N_{\downarrow}}
{N_{\uparrow}+N_{\downarrow}}
=\frac{E_{FN} - U}{2E_{FN}}$]
and
$k_{N,\uparrow} =
\mid \mbox{\boldmath $k$}_{N,\uparrow} \mid \equiv
\sqrt{\frac{2m}{\hbar^{2}}(E_{FN} + U)}$
[$k_{N,\downarrow} =\mid \mbox{\boldmath $k$}_{N,\downarrow}
\mid
\equiv \sqrt{\frac{2m}{\hbar^{2}}
(E_{FN} - U)}$],
respectively \cite{deJong}.
\par
We assume the quasiparticle injection of up spin
electrons at an angle $\theta_N$ to the interface normal
as shown in Fig.~\ref{fig1}.
Four possible trajectories exist;
they are Andreev reflection (AR), normal reflection (NR),
transmission to superconductor as electron-like
quasiparticles (ELQ),
and transmission as hole-like quasiparticles (HLQ).
The spin direction is conserved for NR but not for AR.
When the superconductor has $d_{{x}^{2}-{y}^{2}}$-wave
symmetry,
the effective pair potentials for ELQ and HLQ  are
given by $\Delta_{+} \equiv \Delta_{0} \cos
2(\theta_{S}-\beta)$
and $\Delta_{-} \equiv \Delta_{0} \cos 2(\theta_{S}+\beta)$,
respectively,
where $\beta$ is the angle between $a$-axis of the crystal
and the
interface normal.
Results for various pairing symmetries are obtained by
setting
proper values to $\Delta_{+}$ and $\Delta_{-}$
similarly to the previous formulas \cite{Tanaka,Kashiwaya2}.
The wave vectors of ELQ and HLQ are approximated by
$k_{S}=\mid \mbox{\boldmath $k$}_{S}
\mid\approx
\sqrt{\frac{2m E_{FS}}{\hbar^{2}}}$
following the model by Andreev \cite{Andreev}.
Since translational symmetry  holds along the $y$-axis
direction,
the momentum components of all trajectories are conserved
(${k}_{N,\uparrow}\sin\theta_{N}=
k_{N,\downarrow} \sin\theta_{A}
=k_{S} \sin\theta_{S}$) \cite{Zuc2}.
Note that $\theta_{N}$ is not equal to $\theta_{A}$ except
when $U=0$,
which means retro-reflectivity of AR is broken.
Such novel behavior is a consequence of the
fact that in the presence of an exchange field
the BCS paring is formed not strictly between states of
equal but
opposite $k$-vectors,
the so-called Fulde-Ferrell effect \cite{Fulde}.
The wave-function
in the ferromagnet ($x<0$)
for up [down] spin with injection angle $\theta_{N}$ is
described by
\begin{equation}
\left(\begin{array}{c} u(\mbox{\boldmath $x$},\theta_{N})\\
v(\mbox{\boldmath $x$},\theta_{N})\end{array} \right)
=
e^{i\mbox{\boldmath $k$}_{N,\uparrow[\downarrow]}
\mbox{\boldmath $x$}}
\left(
\begin{array}{c}1\\
0\end{array}
\right)
+a_{\uparrow[\downarrow]}(E,\theta_{N})
e^{i\mbox{\boldmath $k''$}_{N,\downarrow[\uparrow]}
\mbox{\boldmath $x$}}
\left(\begin{array}{c} 0\\
1\end{array} \right)
+ b_{\uparrow[\downarrow]}(E,\theta_{N})
e^{i\mbox{\boldmath $k'$}_{N,\uparrow[\downarrow]}
\mbox{\boldmath $x$}}
\left(
\begin{array}{c}
1\\
0
\end{array} \right),
\end{equation}
where the signs of the $x$-components of
$\mbox{\boldmath $k$}_{N,\uparrow[\downarrow]}$
and  $\mbox{\boldmath $k'$}_{N,\uparrow[\downarrow]}$ are
the reversed of each other.
The reflection probabilities of the two processes
are obtained by solving Eq.~(\ref{BdG})
and by connecting the wave-function and its derivative
at $x=0$.
\par
Next,
we will simply explain the Fermi surface effect
by assuming up spin injection.
Various kinds of reflection process are expected
depending on the values of $E_{FN}$, $E_{S}$ and $U$.
For example,  when
$k_{S}<k_{N,\uparrow}$,
total reflection
($\mid b_{\uparrow[\downarrow]}(E,\theta_{N})\mid^2=1$)
occurs when
$\theta_N >
\sin^{-1}(k_{S}/k_{N,\uparrow})\equiv\theta_{c1}$
\cite{Kashiwaya2,Zutic}.
In this case,  the net currents of the spin and the charge
from the ferromagnet to the superconductor vanish.
On the other hand, when
$k_{N,\downarrow}<k_{S}<k_{N,\uparrow}$,
the $x$-component of wave-vector in AR process
($\sqrt{k^{2}_{N,\downarrow}-k^{2}_{S}\sin^{2}\theta_{S}}$)
becomes purely imaginary
for
$\theta_{c1}>\theta_{N}>
\sin^{-1}(k_{N,\downarrow}/k_{N,\uparrow})\equiv\theta_{c2}$.
In this case,
although transmitted quasiparticles from ferromagnet to
superconductor
do propagate,
the Andreev reflected quasiparticles do not propagate (VAR
process).
A finite amplitude of the evanescent AR process still
exists ($\mid a_{\uparrow}(E,\theta_{N})\mid^2>0$)
and the net currents of the spin and the charge
from the ferromagnet to the superconductor do not vanish.
It is easy to check the conservation laws
for the charge, the excitation,
and the spin on the VAR process
following the method presented in Ref.~\cite{BTK}.
The existence of the VAR process has not been
treated in the one-dimensional model \cite{deJong}
because it is a
peculiar feature of a two or three dimensional F/S
interface.
\par
The conductance of the junctions are obtained by extending
previous formula to include the effect of spin
\cite{BTK,deJong,Zutic}.
In the following, consider a situation where
$k_{N,\downarrow}<k_{S}<k_{N,\uparrow}$.
To analyze the transport properties of an $F/I/S$ junction,
two kinds of conductance spectrum are introduced.
The conductance for the charge current is
defined by the charge flow induced
by the up [down] spin quasiparticle injection and is given
by
\begin{equation}
\label{chdef}
{\hat \sigma}_{q,\uparrow[\downarrow]}(E,\theta_{N})
\equiv
{\rm Re}
\left[ 1 + \frac{\lambda_{2}}
{\lambda_{1}}
\mid a_{\uparrow[\downarrow]}(E,\theta_{N})\mid^2-
\mid b_{\uparrow[\downarrow]}(E,\theta_{N})\mid^2
\right]
\end{equation}
(for $0<\mid\theta_{N}\mid<\theta_{c2}$)
\begin{equation}
\label{chdef2}
=
\frac{4 \lambda_{1}
\left[ 4 \lambda_{2} \mid{\hat \Gamma}_{+}\mid^2+
(1+\lambda_{2})^2+Z_{\downarrow[\uparrow]}^2-\mid{\hat
\Gamma}_{+}
{\hat \Gamma}_{-}\mid^2
\{(1-\lambda_{2})^2+Z_{\downarrow[\uparrow]}^2\}
\right] }
{\mid
(1+\lambda_{1}+iZ_{\uparrow[\downarrow]})(1+\lambda_{2}-iZ_{\downarrow
[\uparrow]})
-(1-\lambda_{1}-iZ_{\uparrow[\downarrow]})(1-\lambda_{2}
+iZ_{\downarrow[\uparrow]})
{\hat \Gamma}_{+} {\hat \Gamma}_{-}\mid^2} ,\\
\end{equation}
(for $\theta_{c2}<\mid\theta_{N}\mid<\theta_{c1}$)
\begin{equation}
\label{chdef3}
=
\frac{4\lambda_{1}(1-\mid{\hat \Gamma}_{+}{\hat
\Gamma}_{-}\mid^2)
\{1+(-\kappa_{2}+Z_{\downarrow})^2\} }
{\mid
(1+\lambda_{1}+iZ_{\uparrow})\{1-i(\kappa_{2}+Z_{\downarrow})\}
-(1-\lambda_{1}-iZ_{\uparrow})\{1+i(\kappa_{2}+iZ_{\downarrow})\}
{\hat \Gamma}_{+} {\hat \Gamma}_{-}\mid^2} ,\\
\end{equation}
(for $\theta_{c1}<\mid\theta_{N}\mid<\pi/2$)
\par
=0,
\par
\noindent
where
\[
Z_{\uparrow[\downarrow]}=\frac{Z_{0,\uparrow[\downarrow]}}{\cos\theta_{S}},
\ \
Z_{0,\uparrow[\downarrow]}=\frac{2m({\hat V}_{0}-[+]{\hat
U}_{B})}{\hbar^{2}
k_{S}},\ \
\]
\[
{\hat \Gamma}_{\pm}=\Gamma_{\pm} \exp({\mp i\phi_{\pm}}), \
\
\exp{i\phi_{\pm}}=\frac{\Delta_{\pm}}{\mid \Delta_{\pm}
\mid}, \ \
\Gamma_{\pm}=\frac{E-\sqrt{E^2-\mid \Delta_{\pm} \mid ^2}}
{\mid \Delta_{\pm} \mid},
\]
\[
\lambda_{1}=\frac{k_{N,\uparrow[\downarrow]}\cos\theta_{N}}{k_{S}\cos\theta_
{S}},\ \
\lambda_{2}=\frac{k_{N,\downarrow[\uparrow]}\cos\theta_{A}}{k_{S}\cos\theta_
{S}},\ \
\kappa_{2}=i\lambda_{2}=
\frac{\sqrt{k^{2}_{S}\sin^{2}\theta_{S}-k^{2}_{N,\downarrow}}}
{k_{S}\cos\theta_{S}}.
\]
The conductance for the spin current is defined by
the spin imbalance induced by
the up [down] spin quasiparticle injection,
\begin{equation}
\label{spdef}
{\hat \sigma}_{s,\uparrow[\downarrow]}(E,\theta_{N})\equiv
{\rm Re}
\left[ 1
- \frac{\lambda_{2}}
{\lambda_{1}}
\mid a_{\uparrow[\downarrow]}(E,\theta_{N})\mid^2-
\mid b_{\uparrow[\downarrow]}(E,\theta_{N})\mid^2
\right]
\end{equation}
(for $0<\mid\theta_{N}\mid<\theta_{c2}$)
\begin{equation}
\label{spdef2}
=\frac{4 \lambda_{1}
\left[
-4 \lambda_{2} \mid\Gamma_{+}\mid^2+
(1+\lambda_{2})^2+Z_{\downarrow[\uparrow]}^2-\mid{\hat
\Gamma}_{+}
{\hat \Gamma}_{-}\mid^2
\{(1-\lambda_{2})^2+Z_{\downarrow[\uparrow]}^2\}
\right]}
{\mid (1+\lambda_{1}+iZ_{\uparrow[\downarrow]})
(1+\lambda_{2}-iZ_{\downarrow[\uparrow]})
-(1-\lambda_{1}-iZ_{\uparrow[\downarrow]})(1-\lambda_{2}
+iZ_{\downarrow[\uparrow]})
{\hat \Gamma}_{+} {\hat \Gamma}_{-} \mid^2} ,
\end{equation}
(for $\theta_{c2}<\mid\theta_{N}\mid<\theta_{c1}$)
\begin{equation}
\label{spdef3}
=
\frac{4\lambda_{1}(1-\mid{\hat \Gamma}_{+}{\hat
\Gamma}_{-}\mid^2)
\{1+(-\kappa_{2}+Z_{\downarrow})^2\} }
{\mid
(1+\lambda_{1}+iZ_{\uparrow})\{1-i(\kappa_{2}+Z_{\downarrow})\}
-(1-\lambda_{1}-iZ_{\uparrow})\{1+i(\kappa_{2}+iZ_{\downarrow})\}
{\hat \Gamma}_{+} {\hat \Gamma}_{-}\mid^2} ,\\
\end{equation}
(for $\theta_{c1}<\mid\theta_{N}\mid<\pi/2$)
\par
=0.
\par
\noindent
The Andreev reflected quasiparticles  positively contribute
to
the charge current, but since their spins are reversed, they
have
negative contribution to the
spin current.
Second terms in r.h.s. of Eqs.~(\ref{chdef}) and
(\ref{spdef})
do not have finite contribution on net current
in the VAR process,
since the corresponding $\lambda_2$ is purely imaginary.
The normalized total conductance spectra for the charge
current
$\sigma_{q}(E)$
and the spin current $\sigma_{s}(E)$ are given by
\begin{equation}
\sigma_{q}(E)=\sigma_{q,\uparrow}(E)+\sigma_{q,\downarrow}(E),
\end{equation}
\begin{equation}
\sigma_{q,\uparrow[\downarrow]}(E)=
\frac{1}{R_N}
\int_{-\pi/2}^{\pi/2} d\theta_{N} \cos\theta_{N}
{\hat \sigma}_{q,\uparrow[\downarrow]}(E,\theta_{N})
P_{\uparrow[\downarrow]} k_{F,\uparrow[\downarrow]},
\end{equation}
\begin{equation}
\sigma_{s}(E)=\sigma_{s,\uparrow}(E)-\sigma_{s,\downarrow}(E),
\end{equation}
\begin{equation}
\sigma_{s}(E)=\frac{1}{R_N}
\int_{-\pi/2}^{\pi/2} d\theta_{N} \cos\theta_{N}
{\hat \sigma}_{s,\uparrow[\downarrow]}(E,\theta_{N})
P_{\uparrow[\downarrow]} k_{F,\uparrow[\downarrow]},
\end{equation}
where
\begin{equation}
R_N=\int_{-\pi/2}^{\pi/2} d\theta_{N} \cos\theta_{N}
\left[{\hat \sigma}_{N,\uparrow}(\theta_{N}) P_{\uparrow}
k_{F,\uparrow}
+{\hat \sigma}_{N,\downarrow}(\theta_{N})
P_{\downarrow} k_{F,\downarrow}
\right],
\end{equation}
\[
{\hat
\sigma}_{N,\uparrow[\downarrow]}(\theta_{N})=\frac{4\lambda_{1}}
{\mid 1+\lambda_{1}+iZ_{\uparrow[\downarrow]} \mid^{2}}.
\]
In the above, $R_N$, $\sigma_{q,\uparrow[\downarrow]}(E)$
and
$\sigma_{s,\uparrow[\downarrow]}(E)$
correspond to the conductance when the superconductor
is in the normal state and the spin-resolved normalized
conductance
spectra for charge and spin,
respectively.
The net polarization $J_{p}(eV)$ as a function of the
bias voltage $V$ is give by
\begin{equation}
J_{p}(eV)=\frac{\int_{-\infty}^{\infty} dE
\sigma_s(E) \{f(E-eV)-f(E)\}}
{\int_{-\infty}^{\infty} dE \sigma_q(E) \{f(E-eV)-f(E)\}},
\end{equation}
where $f(E)$ is the Fermi distribution function.
Since the convolution with $f(E)$ gives only a smearing
effect
in the conductance spectra,
the temperature is set to zero in the following discussions.
\par
In the above formulation, we have neglected
the self-consistency of the pair potential
in order to get analytical formulas \cite{Barash}.
However, the present  formula is easily extended to
include this effect simply by replacing $\Gamma_{\pm}$
with $\Gamma_{\pm}(x)\mid_{x=0}$, where $\Gamma_{\pm}(x)$
follows the Ricatti equations described by
\begin{equation}
\frac{d}{dx} {\hat \Gamma}_{+}
(x)=
\frac{1}{i \hbar^2 \ {k}_{F}\cos\theta_{S} }
\left[ -{\Delta}_{+}(x)
{\hat \Gamma}_{+}^{2}(x)  -{\Delta}^{*}_{+}(x)
+ 2 E{\hat \Gamma}_{+}(x) \right],
\end{equation}
\begin{equation}
\frac{d}{dx} {\hat \Gamma}_{-}
(x)=
\frac{ 1}{i \hbar^2 k_{F}\cos\theta_{S} }
\left[ -{\Delta}^{*}_{-}(x) {\hat \Gamma}_{-}^{2}
(x)  -{\Delta}_{-}(x)
+2E {\hat \Gamma}_{-}(x) \right],
\end{equation}
Here the spatial dependence of the pair potential
is assumed
as $\Delta_{\pm}(x)$ (functions of $x$).
\par
The  most important differences in the present formula from
previous ones are; i) a novel formula for
the non-linear spin current,
ii) a capability to treat
the ferromagnetic insulator effects based on the scattering
method,
iii) the introduction of the breakdown in the
retro-reflectivity
of the AR process
and consequently the vanishing of the propagating AR
(VAR process).
In particular, the concept of the VAR process is a new
physical process
presented in this paper.
If we would not accept the existence of this process,
the total reflection independent of $E$ is naively expected.
Since finite transmission is possible
in this angle region ($\theta_{c2}<\mid \theta
\mid<\theta_{c1}$) above $T_{c}$,
this total reflection would induce
a sudden decrease of the conductance
just below $T_{c}$
for highly polarized ferromagnets junctions.
As far as we know, no trends for such effect has been
reported thus far.
This fact may be the direct evidence
for the existence of the VAR process.
The VAR process is shown to have an important role
on the Josephson current in
superconductor/ferromagnet/superconductor
junctions,
because the evanescent wave carries a net Josephson current
in this configuration \cite{TanaSFS}.
Note that the suppression mechanism of the AR process
presented here is essentially different from that discussed
in one-dimensional model where
the contribution of the AR to the net current is
simply governed by the ratio
$k_{N,\downarrow}/k_{N,\uparrow}$ \cite{deJong}.
\par
\section{Results}
\label{sec3}
\subsection{Effects of polarization}
\label{sec3-1}
In this subsection, to reveal the influence of the
polarization
on the tunneling conductance spectra,
we assume F/I/S junction by setting ${\hat U}_B=0$
($Z_{0,\uparrow}=Z_{0,\downarrow}
\equiv Z_{0}$).
At first, let us discuss several analytical results
obtained from above formulation
in order to check the validity of the formula.
When $U=0$, the ferromagnet reduces to a normal
metal, and as expected $\sigma_{q}(E)$ reproduces the
results of
Ref.~\cite{Tanaka,Kashiwaya2}, and $\sigma_{s}(E)$ vanishes.
For half-metallic ferromagnets ($U=E_{FN}$),
the Fermi-surface for the down spins has shrunk to zero.
In this case, the VAR process occurs for all $\theta_{N}$.
Under the condition of VAR, ${\hat \sigma}_{q}(E,\theta_{N})
={\hat \sigma}_{s}(E,\theta_{N})$ applies,
which corresponds to the fact that
the tunneling current is completely spin-polarized.
Furthermore,
the conductance spectra in the energy gap
($E<\mid\Delta_{+}\mid$, $E<\mid\Delta_{-}\mid$)
become completely zero
[$\sigma_{q}(E,\theta_{N})=\sigma_{s}(E,\theta_{N})=0$].
In the tunneling limit ($H \rightarrow \infty$)
and in the absence of VAR,
${\hat \sigma}_{q,\uparrow[\downarrow]}(E, \theta)$ gives
the
angle resolved surface
DOS of an isolated superconductor.
Then $\sigma_{q}(E)$
converges to the surface DOS weighted by
the tunneling probability distribution \cite{Kashiwaya2}.
At this limit, we can reproduce a well-known result that
the ratio of the peak heights in the spin-resolved spectra
directly
reflect the polarization in the ferromagnet \cite{Tedrow}.
On the other hand, ${\sigma}_{s,\uparrow[\downarrow]}(E,
\theta)$
reduces to a function similar to the surface DOS,
but where the divergence at the energy levels of the surface
bound states
is missing.
\par
Next, the calculated results based on
above formula are presented for $d_{{x}^{2}-{y}^{2}}$-wave
superconductors.
In the following, we assume $E_{FN}=E_{FS}$.
Figures~\ref{fig2} and ~\ref{fig3} show the
conductance spectra of charge current for
the transparent limit ($Z_{0}=0$, $\beta=0$)
and high-barrier case ($Z_{0}=5$, $\beta=\pi/4$)
as the function of exchange interaction $X(\equiv
U/E_{FN})$.
For $X=0$, results in Ref.~\cite{Tanaka} are reproduced.
However, as $X$ increases, the conductance inside the gap
($\mid E\mid <\Delta_{0}$) is largely
reduced for both cases.
Especially, the ZBCP disappears for the  half-metallic
ferromagnet
case.
Since the spin-polarization has such a  drastic influence on
the ZBCP,
the height of ZBCP can be used in principle as a measurement
of
the magnitude of the spin polarization.
Figure~\ref{fig4} shows the difference of
the spin current and the charge current
when $X$=0.85, $Z_{0}=5$ and $\beta=\pi/4$.
It is clear that the ZBCP is not present for the spin
current.
This corresponds to the fact that the charge current
components corresponding
to the ZES are carried by
condensed  Cooper pairs
in the superconductor, and therefore
they do not contribute to the spin imbalance.
As a result, the spin current becomes relatively insensitive
to the orientation of the junctions.
Figure~\ref{fig5} shows the conductance spectra for
the spin current as the function of
spin polarization ($Z_{0}=5$). 
It is clear that the spin current increases as $X$ becomes
larger.
Note that $\sigma_{s}(E)$
is larger than unity around
$E=\Delta_0$ when $X \approx 1$.
This corresponds to the fact that the peak in the DOS
has an influence even for the spin current.
\par
Next, the net polarization $J_{p}(eV)$ is calculated for
$d_{{x}^{2}-{y}^{2}}$-wave superconductors
as a function of the orientation ($\beta$) when $T=0$.
Four lines of Fig.~\ref{fig6} show the results
for various values of barrier parameter
when $eV=2\Delta_0$.
It is clear that the orientational effect
is much smaller compared to the effect of $Z_0$.
In the same figure, results for $s$-wave superconductors
($\Delta_+=\Delta_-=\Delta_0$ independent of $\theta_N$)
are also shown as closed dots.
The large deviations of $d_{{x}^{2}-{y}^{2}}$-wave
from $s$-wave for small values of $Z_0$ are originated
from the distribution of the pair amplitude in $k$-space.
As the barrier parameter becomes larger,
the spin injection efficiency
becomes to be insensitive to the symmetry of the pair
potential.
\par
\subsection{Spin filtering effects and the ZBCP splitting}
\label{sec3-2}
It has been experimentally verified that
a ferromagnetic semiconductor used as the insulator
in tunneling junctions works as a ferromagnetic barrier.
Since the transmission probabilities for up and down spins
are
not equal, an spin-filtering effect is expected
to be realized \cite{Sfilter1,Sfilter2}.
Also it has been theoretically verified that a ferromagnetic
insulator
placed in the vicinity of superconductor induces a
spin-splitting
on the DOS of $s$-wave superconductors \cite{Tokuyasu}.
In the following, we will analyze the influence of the
exchange interaction
existing inside the insulator
on the transport properties
%in the cases of $d_{{x}^{2}-{y}^{2}}$-wave superconductors
based on the formulation described in Sec.~\ref{sec2}.
\par
Figure~\ref{fig7} shows the response of
the conductance spectra $\sigma_{q}(E)$
on the exchange interaction in the insulator
when $X=0$.
ZBCP splittings are obtained for finite exchange amplitude
($U_B$) cases.
As $U_B$ is increased and consequently as the difference
between $Z_{0,\uparrow}$ and $Z_{0,\downarrow}$ becomes
larger,
the amplitude of the splitting  becomes larger
and the two peaks become broader and smaller.
The peaks in the gap disappear when the difference
between $Z_{0,\uparrow}$ and $Z_{0,\downarrow}$
becomes prominent.
To see more clearly these trends, the spin-resolved
conductance spectra
$\sigma_{q,\uparrow[\downarrow]}(E)$ and
$\sigma_{s}(E)$ for
$Z_{0,\uparrow}=2.5$ and $Z_{0,\downarrow}=7.5$ are plotted
in
Fig.~\ref{fig8}.
The spectra for up [down] spins are shifted for lower
[higher]
energy level.
Furthermore, $\sigma_{s}(E)$ becomes finite even
though $X=0$ in the ferromagnet.
In order to check the effect of the polarization,
Fig.~\ref{fig9} shows the response of the charge current
as a function of polarization $X$
for a fixed barrier parameter.
The spin polarization in the ferromagnet induces the
imbalance
of the peak heights, thus the ratio
of the splitted peak heights can be used as a criteria
for the spin-polarization.
\par
These results are interpreted as follows:
i) The peaks corresponding to the up [down] spin
components are shifted because of the
energy gain (loss) during the tunneling process.
ii) Since this energy gain (loss) has $k$-dependence
\cite{Tokuyasu},
the peak becomes broader comparing to the
magnetic-field induced peak splitting (see below).
iii) The amplitude of the peak splitting  depends
on the
genuine barrier amplitude ${\hat V}_0$
as well as the exchange amplitude ${\hat U}_B$.
For example, the splitted peaks merge into a single peak
at the tunneling limit (${\hat V}_0\rightarrow \infty$)
even if ${\hat U}_B$ is kept constant.
iv)
The current corresponding to the ZBCP is
carried by the Cooper pair in the superconductor
as described in the previous subsection.
This corresponds to the fact that the AR process
is the second-lowest order tunneling process
which requires both
up and down spins tunneling.
Hence, as  $Z_{0,\downarrow}$ becomes larger and as the
tunneling probability for down spins are suppressed,
the conductance peaks and the AR process
are rapidly reduced even if $Z_{0,\uparrow}$ is kept zero.
v) The spin current is increased  as ${\hat U}_B$ is raised
from zero
even if $X$ in the ferromagnet is kept at zero
(unpolarized).
This feature directly corresponds to the spin-filtering
effect
that the spin-selective tunneling occurs
due to the presence of the exchange field in the insulator.
\par
Next, various types of the ZBCP splitting
expected for $d$-wave superconductors
and their polarization effects are analyzed.
Mainly two
possibilities other than the ferromagnetic insulator effects
have been proposed for the origins of the ZBCP splitting
on high-$T_c$ superconductor junctions.
One is the Zeeman effect due to an applied magnetic field,
and the other is the inducement of the BTRS
states such as $d_{x^2-y^2}$+$is$-wave.
The conductance spectra
in an applied magnetic field is calculated from above
formula
by simply using the relation
\begin{equation}
\sigma_{q[s]}(E)=\sigma_{q[s],\uparrow}(E-\mu_{B}H)+
\sigma_{q[s],\downarrow}(E+\mu_{B}H),
\end{equation}
where $\mu_B H$  is the Zeeman energy.
Calculated charge conductance spectra 
for $d_{x^2-y^2}$-wave superconductor
as a function of $X$ are shown in Fig.~\ref{fig10}.
The amplitude of the splitting is linear to the applied
field
independent of the barrier heights.
Moreover, since the energy shift induced by 
the magnetic field does not have
$k$-dependence, the broadening of the peaks are not
observed.
The ratio of the splitted peak heights simply reflects the
polarization
in the ferromagnet, which
is consistent with the results by Tedrow and Meservey
\cite{Tedrow}.
On the other hand, $\sigma_{q}(E)$
for $d_{x^2-y^2}$+$is$-wave superconductor is calculated by
setting $\Delta_{\pm}=\Delta_{0} \cos
2(\theta_{S}\mp\beta)+i\Delta_{s}$.
Calculated charge conductance spectra
for various $X$ values are shown in Fig.~\ref{fig11}.
The amplitude of the splitting is almost
equivalent to the amplitude of the $s$-wave component.
The shape of the spectrum without the polarization ($X=0$)
is quite similar to that shown in Fig.~\ref{fig10} ($X=0$).
As $X$ becomes larger,
the heights of the two peaks are reduced, which 
is consistent with that shown in Fig.~\ref{fig3}.
On the other hand,
differently from Figs.~\ref{fig9} and \ref{fig10},
since the peak splitting is not induced
by spin-dependent effects in this case,
the polarization in the ferromagnets
does not yield an imbalance in the peak heights.
Thus the heights of the two peaks are reduced symmetrically.
\par
The responses of the ZBCP on the variation of
the polarization and the applied
magnetic field  are summarized as follows:
i) the peak splitting due to the ferromagnetic insulator and

the Zeeman effect 
are spin dependent. Therefore, the polarization
in the ferromagnet induces the asymmetrical splitting of the
ZBCP.
%Inversely, the BTRS state does not yield the asymmetry.
ii) The amplitude of the peak splitting is linear  to the
applied field
in the case of the  Zeeman effect.
However, it is non-linear in
the cases of  the ferromagnetic insulator effects
\cite{Sfilter2}
and the BTRS states \cite{Fogel}.
In particular, the peak splittings are expected even
in the absence of the applied field for these two cases.
iii) The combination of the BTRS states and the Zeeman
effect induces
an additional peak splitting, that is,  the ZBCP splits
into four peaks. However,  the combination of the Zeeman 
and
the ferromagnetic insulator effects yields two peaks.
\par
The experimental observations of the ZBCP splitting
have been reported for normal metal / high-$T_c$
superconductor 
junctions \cite{Kashiwaya1,Covington,Lesueur,Kash-expt}.
It is a really interesting experiment to observe the same
features
by using ferromagnets/high-$T_c$ superconductor junctions
in order to distinguish the spin-dependent effects from the
BTRS states inducement.
Recently, Sawa, $et. al$. have detected an asymmetric
magnetic field response
in La$_{0.67}$Sr$_{0.33}$MnO$_{3}$ /
YBa$_{2}$Cu$_{3}$O$_{7-\delta}$ junctions
\cite{Sawa}.
The qualitative features on the magnetic field responses
of their junctions are  consistent with F/FI/S
with $d_{x^2-y^2}$-wave explained here.
Detailed comparison between above formulas and their
experiments
is strongly expected.
\par
Finally, a simple proposal is given for a possible device
application 
utilizing the
ferromagnetic insulator effects.
The thickness of the insulator is
the order of 1nm in usual tunneling junctions.
Since the controlling of properties in such a thin layer
requires high technology, as far as we know,
not so many experimental trials have been accomplished thus
far.
However, as shown in this paper, a small change
in the insulator property causes
a drastic change on the transport properties.
Therefore, the  controlling of the barrier properties 
is one of the most
promising methods to create new functional devices.
For example, consider a F/FI/S junction with a
$d_{x^2-y^2}$-wave superconductor
($\beta=\pi/4$).
The sharp ZBCP
is drastically modified as the difference between
$Z_{\uparrow}$ and $Z_{\downarrow}$ becomes larger
as shown in Fig.~\ref{fig7}.
This means that, for a fixed bias voltage, a large response
in current
is expected due to a small variation in the exchange
interaction
in the insulator.
This response is applicable for the high-sensitive
magnetization measurement of
a thin insulating film by inserting the film into a
junction as a tunneling barrier.
If the exchange interaction is sensitive to the external
field,
this effect can be used as a magnetic sensor.
Alternatively, if the magnetization of the insulator shows a
hysteresis
on the external field variation,
a memory function can be realized.
The current gain of the junction as
a function of the external field
is largely enhanced by using a superconductor /ferromagnetic
insulator
/superconductor junction with $d$-wave,
because  negative-conductance regions are expected
just beside the ZBCP
in this configuration \cite{KashiwayaSIS,YoshidaAC}.
Differently from conventional superconducting memories
based on a flux-quantum logic,
a large-scale integration circuit
may be possible based on the present principle.
\par
\section{Summary}
\label{sum}
In this paper,
the conductance spectra for the charge and the spin currents
under the influence of the exchange interaction
have been calculated based on the scattering method.
The influence of the spin polarization on the transport
properties has been clarified.
It is shown that the retro-reflectivity of the standard
Andreev reflection process is broken in the presence of an
exchange field and that the surface bound states due to
superconducting
pair potentials do not contribute
to the spin current.
Next, the ferromagnetic insulator including the
spin-filtering effect are analyzed.
It is shown that the spin-polarization gives
asymmetric peak splitting.
Moreover, various features in the splitting of ZBCP
due to the ferromagnetic insulator, the Zeeman splitting,
and
the  BTRS states effects are analyzed in detail.
It is shown that the
spin-polarized tunneling
gives quite important information
to identify the origin of the ZBCP splitting.
By comparing the present analysis with experimental data,
we expect that the mechanism of the peak splitting in
high-$T_c$ superconductors will be well identified.
In the present model,  we have neglected the effects of
spin-orbit scattering \cite{Fulde2} and the non-equilibrium
properties of superconductors \cite{Clarke,Tinkham}.
Inclusion of these effects would be necessary for a complete
theory.
The formulation for triplet superconductors will be
presented in
another publication \cite{Yoshida}.
\par
\section{Acknowledgments}
We would like to thank M. Koyanagi, K. Kajimura, M.
Yamashiro,
J. Inoue,  A. Sawa, and D. Worledge for
fruitful discussions.
This work has been partially supported by the
the U. S. Airforce Office of Scientific Research.

\begin{figure}
\caption{
Schematic illustration of the
elastic reflection of quasiparticles in the F/I/S junction.
For all trajectories, momenta
parallel to the interface are conserved.
This means that
the retro-reflection property of Andreev process
is lost  due to the exchange interaction.
In the figure, the anisotropic pair potential of
$d_{{x}^{2}-{y}^{2}}$-wave symmetry
is also shown.}
\label{fig1}
\end{figure}

\begin{figure}
\caption{
The normalized conductance spectra for the charge current 
$\sigma_{q}(E)$
as the function of $X\equiv U/E_{FN}$
with  $\beta=0$ and $Z_{0}=0$
(the transparent limit).
As $X$ becomes larger, the peak around zero-bias level is
largely suppressed.
}
\label{fig2}
\end{figure}
\begin{figure}
\caption{
The normalized conductance spectra
for the charge current $\sigma_{q}(E)$
as the function of $X$
with $\beta=\pi/4$ and $Z_{0}=5$.
As $X$ becomes larger,
the height of the ZBCP is largely reduced.
}
\label{fig3}
\end{figure}
\begin{figure}
\caption{
The comparison between the normalized
conductance spectrum for the charge current $\sigma_{q}(E)$
and that for the spin current $\sigma_{s}(E)$ for $X=0.7$,
 $\beta=\pi/4$ and $Z_{0}=5$.
Since the ZBCP originates from the current carried by the
surface bound states,
the peak disappears for the spin current.
}
\label{fig4}
\end{figure}

\begin{figure}
\caption{
The normalized conductance spectra
for the spin current  $\sigma_{s}(E)$
as the function of $X$
with $\beta=0$ and $Z_{0}=5$.
As $X$ becomes larger,
the spin current is increased.
Note that the peak at $E=\Delta_d$ is larger than unity
when $X$ is close to one.
}
\label{fig5}
\end{figure}

\begin{figure}
\caption{
Orientational dependencies
of $J_p(E)$ for $d_{{x}^{2}-{y}^{2}}$-wave
superconductors for $E=2\Delta_0$
and $X=0.7$ are plotted for various $Z_0$
values.
Closed dots in the figures correspond to those for
$s$-wave superconductors.
The large deviations of $d_{{x}^{2}-{y}^{2}}$-wave
from $s$-wave for small values of $Z_0$ are originated
from the distribution of the pair amplitude in $k$-space.
}
\label{fig6}
\end{figure}

\begin{figure}
\caption{
The effects of ferromagnetic insulator
on the charge current for $X=0$ and $\beta=\pi/4$.
When $Z_{0,\uparrow}=5$ and $Z_{0,\downarrow}=5$,
a large ZBCP exists.
The  difference in $Z_{0,\uparrow}$ and $Z_{0,\downarrow}$
induces the peak splitting ($Z_{0,\uparrow}=3$ and
$Z_{0,\downarrow}=7$).
As the difference becomes larger,
the ZBCP split into two peaks and the
amplitude of the splitting  becomes larger
and the peaks become smaller and broader
($Z_{0,\uparrow}=2$ and $Z_{0,\downarrow}=8$).
Finally, the peaks in the gap disappear as the difference
becomes prominent ($Z_{0,\uparrow}=0$ and
$Z_{0,\downarrow}=10$).
}
\label{fig7}
\end{figure}
\begin{figure}
\caption{
Four types of normalized conductance for
$Z_{0,\uparrow}=2$, $Z_{0,\downarrow}=8$,
$X=0$ and $\beta=\pi/4$.
The normalized charge current $\sigma_q(E)$
has splitted peaks.
The peak of the lower energy
and higher energy are originated from
the up spin component $\sigma_{q,\uparrow}(E)$
and down spin component $\sigma_{q,\downarrow}(E)$.
These peaks do not appear in the spin current conductance
$\sigma_s(E)$.
}
\label{fig8}
\end{figure}

\begin{figure}
\caption{
The normalized conductance spectra  $\sigma_{q}(E)$
as a function of $X$ for
$Z_{0,\uparrow}=2.5$, $Z_{0,\downarrow}=7.5$
and $\beta=\pi/4$.
As $X$ becomes larger, the higher energy peak becomes
smaller.
Thus, the polarization can be estimated from the ratio
of two peak heights.}
\label{fig9}
\end{figure}

\begin{figure}
\caption{
The normalized conductance spectra  $\sigma_{q}(E)$
in an applied magnetic field ($\mu_gH/\Delta_0=0.15$)
as a function of $X$ for
$Z_{0,\uparrow}=5$, $Z_{0,\downarrow}=5$,
and $\beta=\pi/4$.
As $X$ becomes larger, the peak with higher energy
is largely reduced.
}
\label{fig10}

\end{figure}
\begin{figure}
\caption{
The normalized conductance spectra  $\sigma_{q}(E)$
for the BTRS states ($\Delta_s/\Delta_0=0.15$)
as a function of $X$ for
$Z_{0,\uparrow}=5$, $Z_{0,\downarrow}=5$
and $\beta=\pi/4$.
As $X$ becomes larger, the heights of the two peaks are
reduced symmetrically.
}
\label{fig11}
\end{figure}


\begin{thebibliography}{10}
\bibitem{Beasley}
Although the present treatment of a
$d$-wave superconductor is valid for
over-doped high-$T_c$ superconductors,
a different treatment may be required for
under-doped cases.
\end{thebibliography}
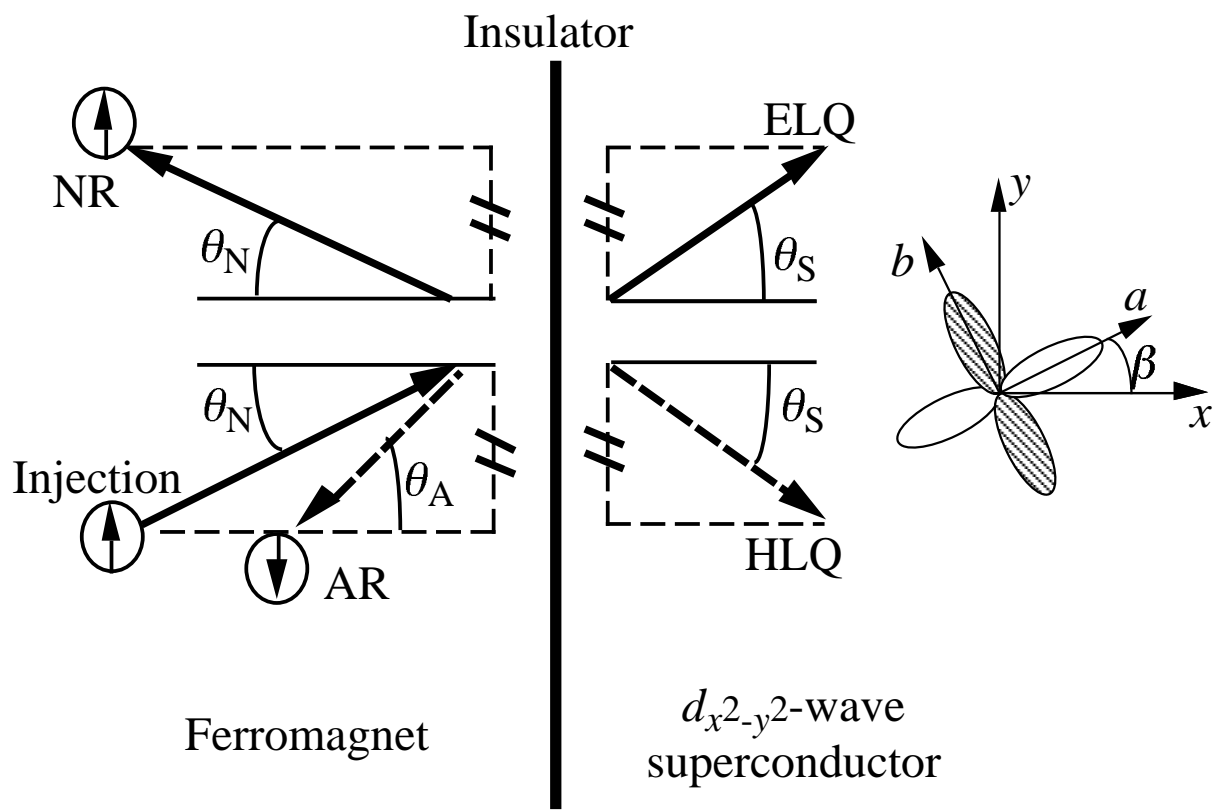
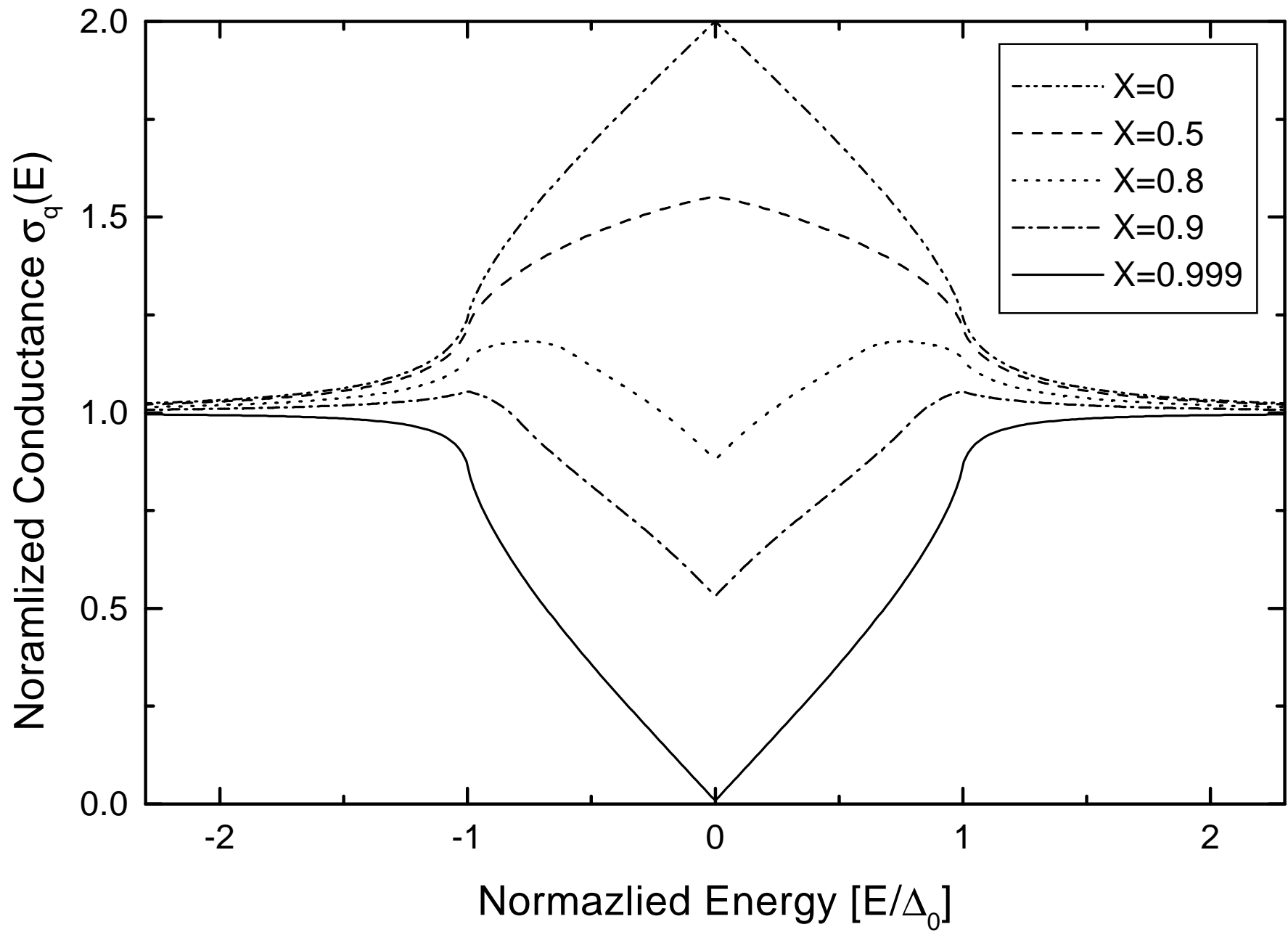
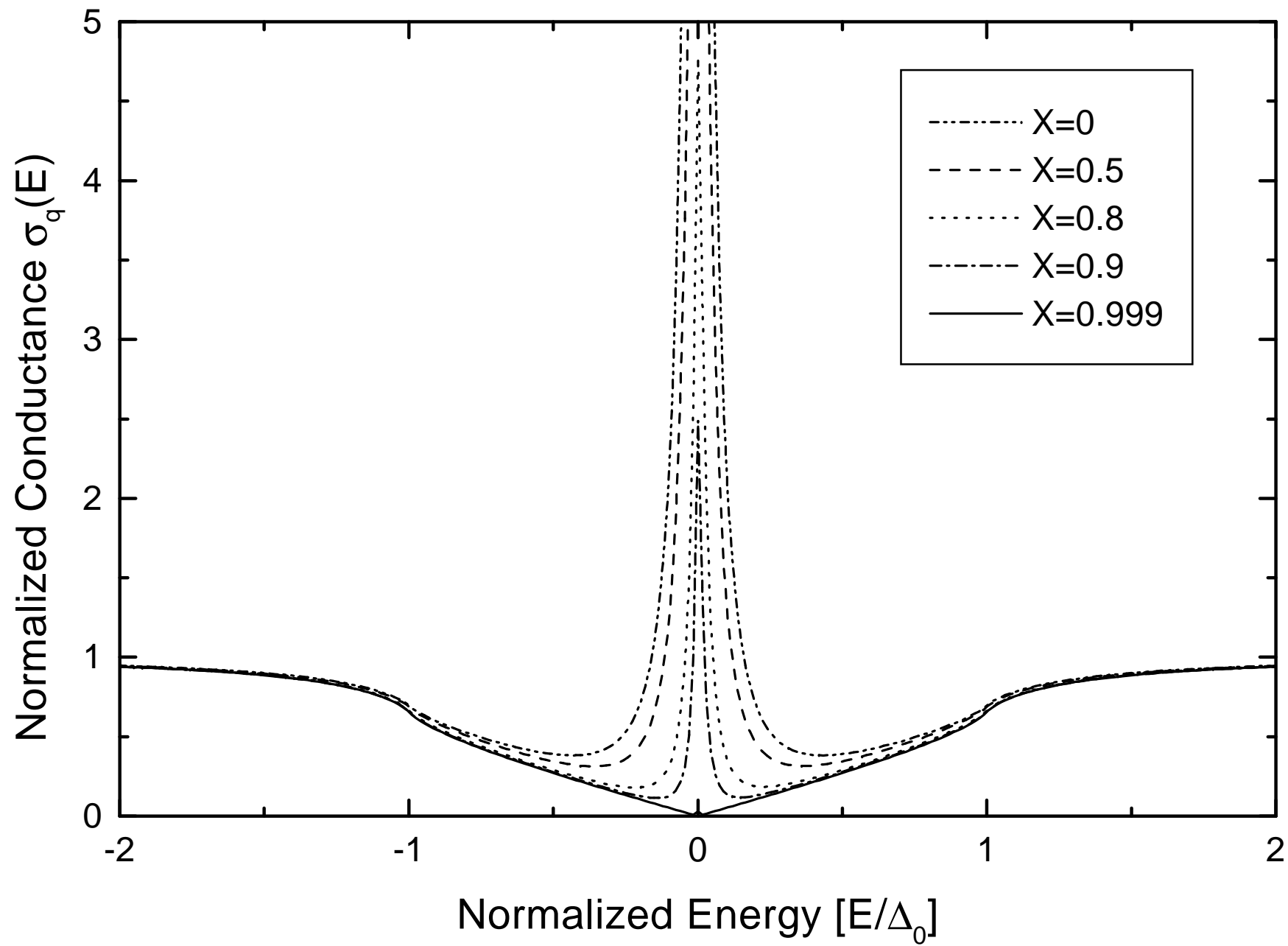
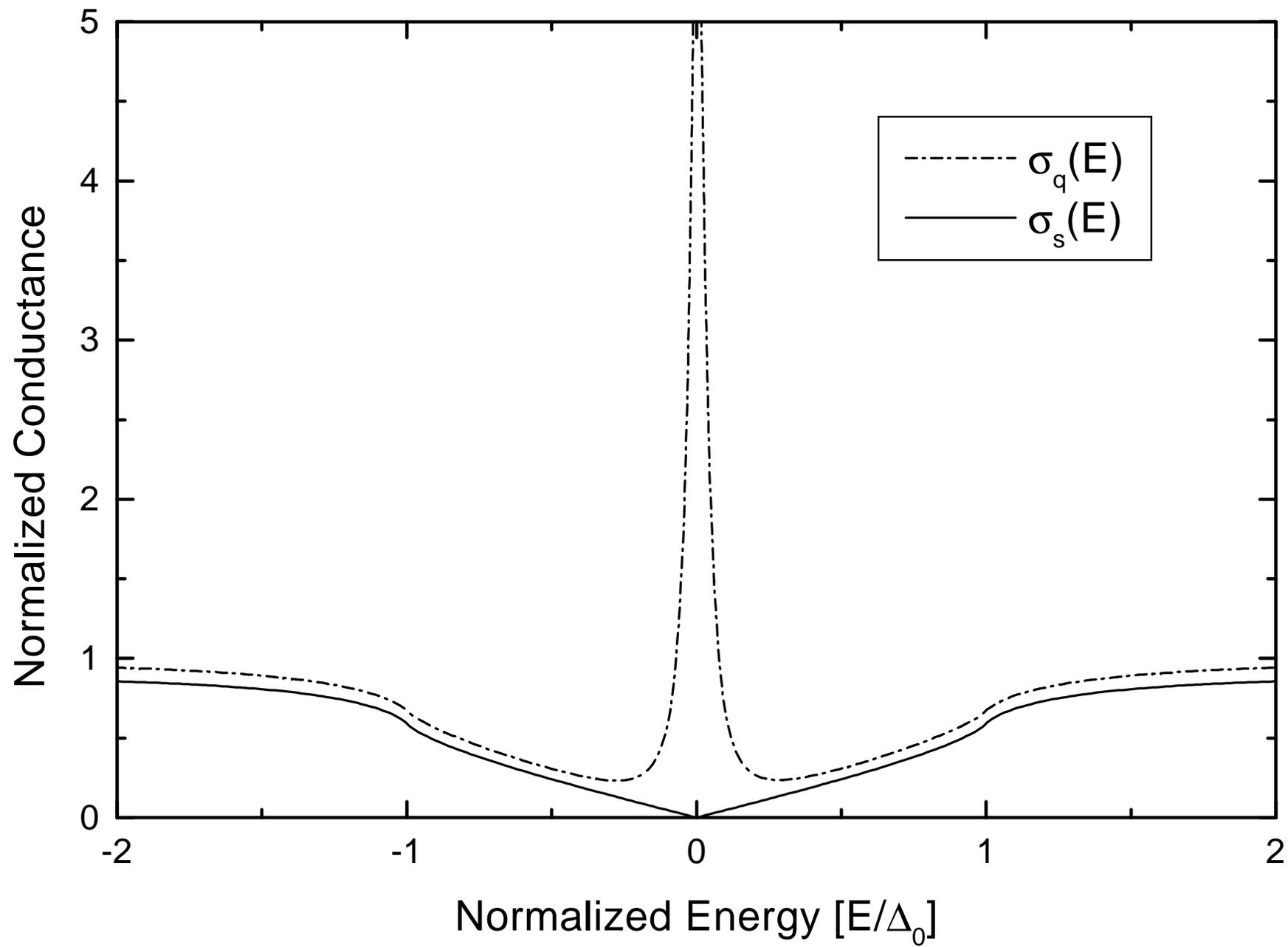
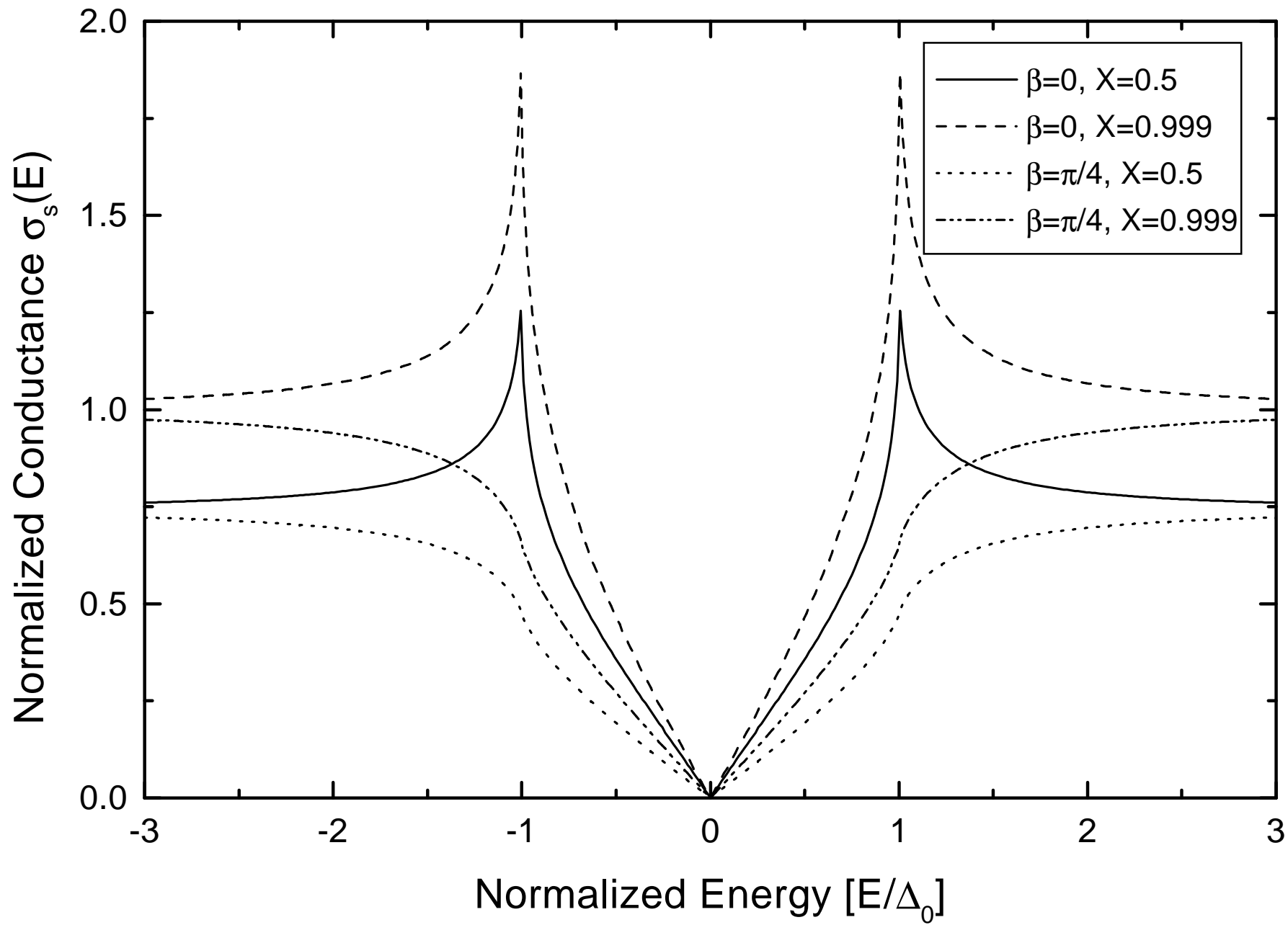
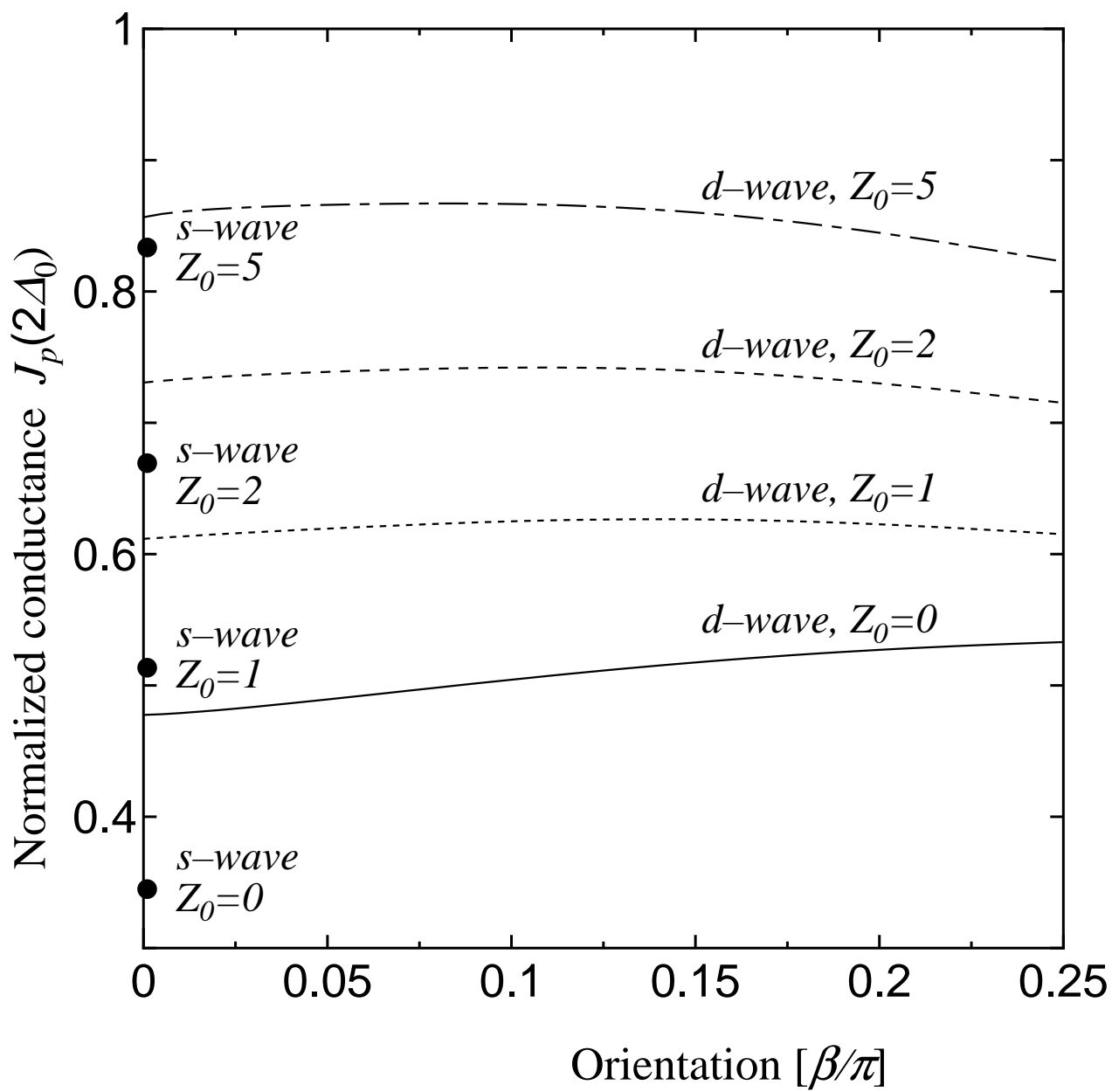
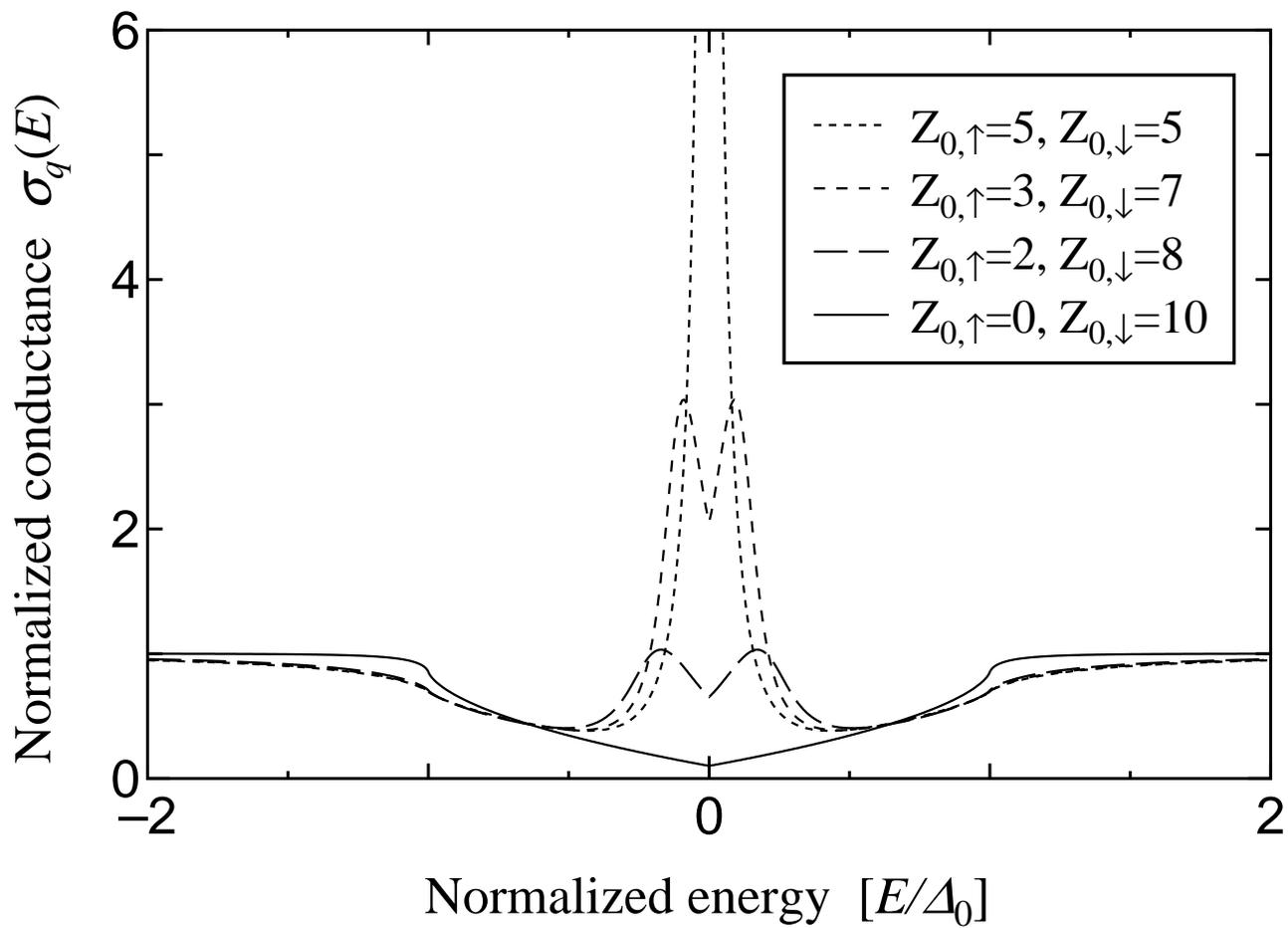
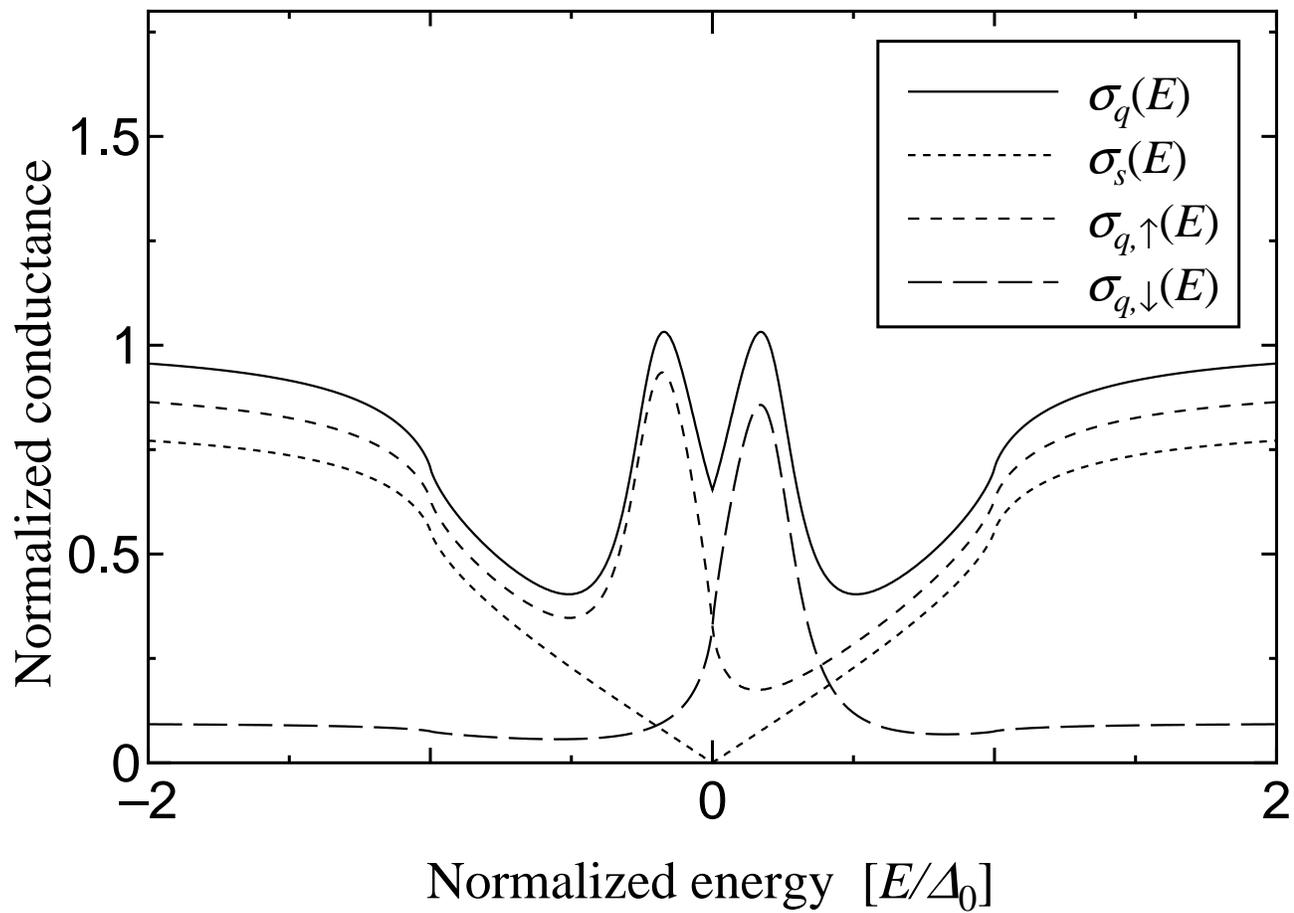
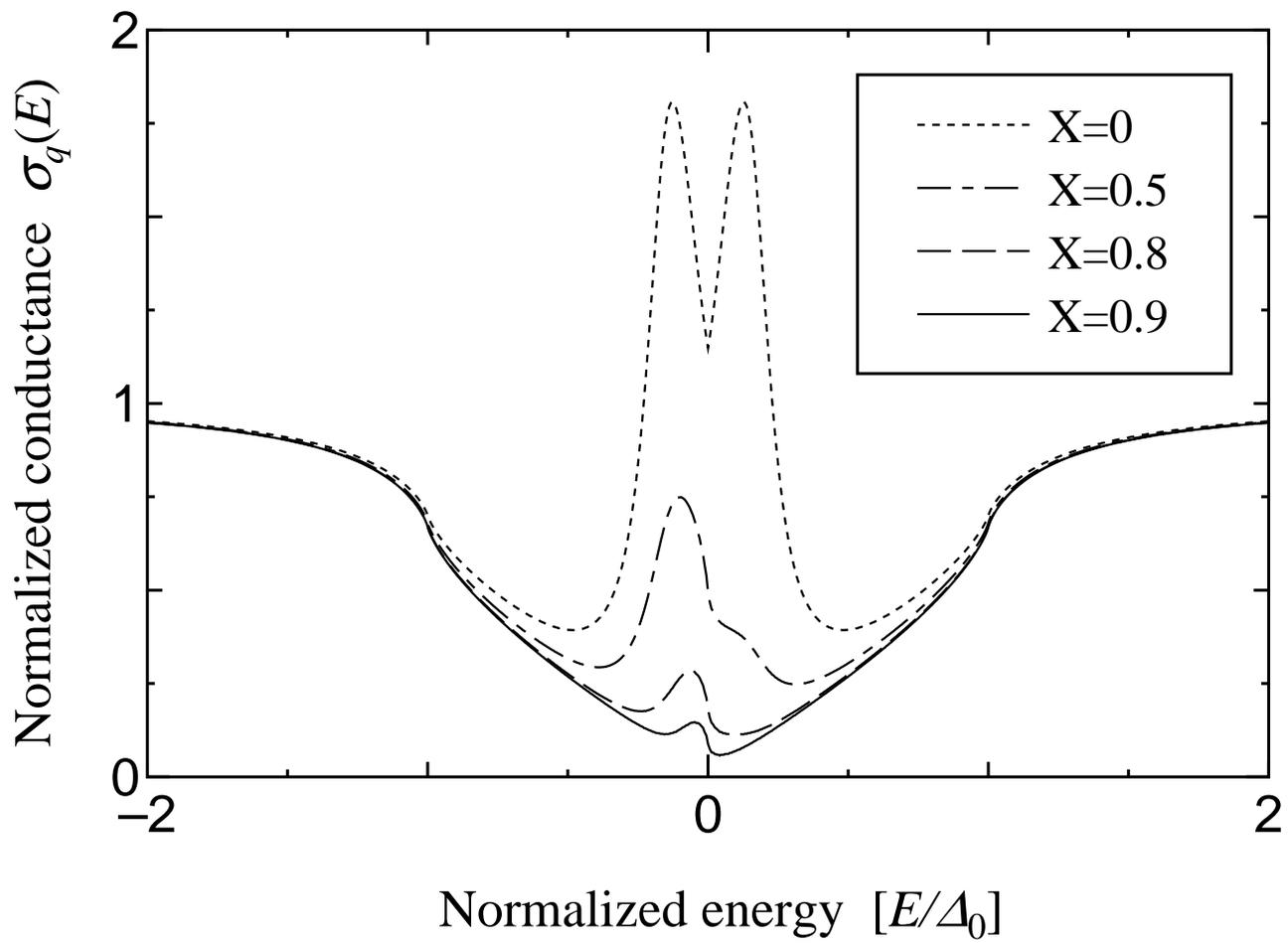
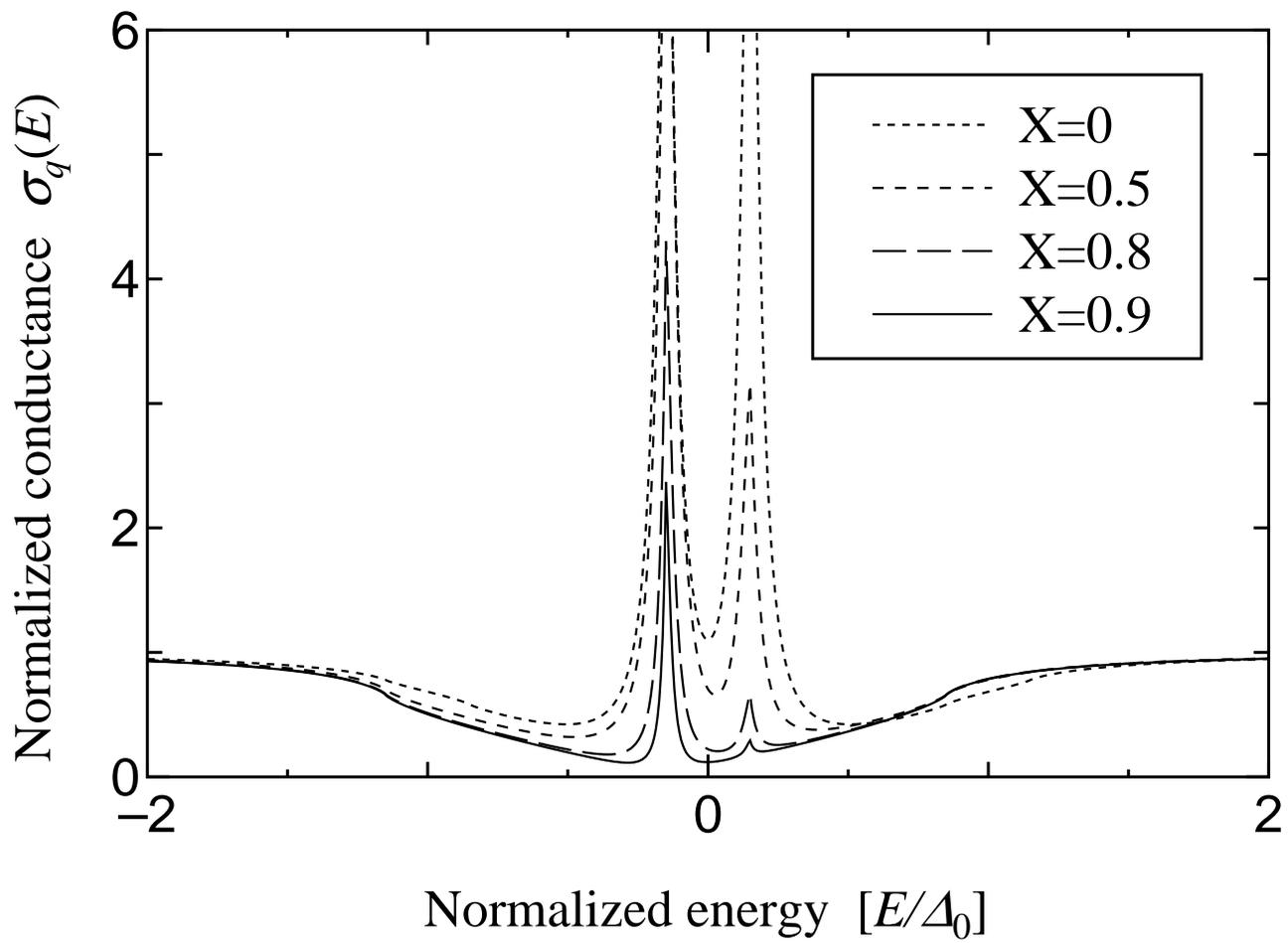
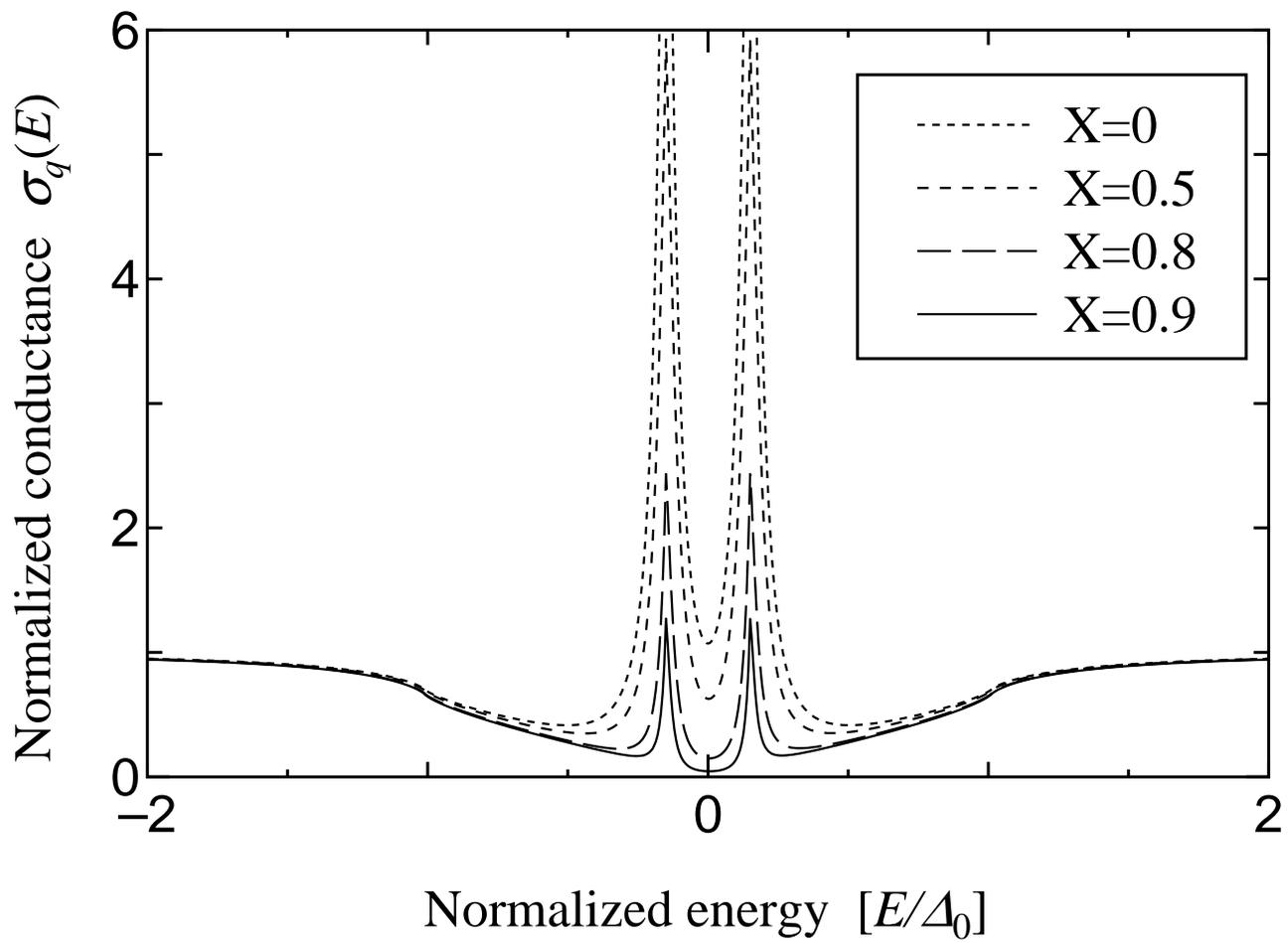
\end{document}